\documentclass[12pt]{iopart}
\usepackage{graphicx}
\begin{document}
\title[Constraining the $0\nu2\beta$ matrix elements by nuclear structure observables]{Constraining the $0\nu2\beta$ matrix elements by                                           nuclear structure observables}
\author{S.J. Freeman$^1$ and J.P. Schiffer$^2$}
\address{$^1$ School of Physics and Astronomy, The University of Manchester, Manchester, M13 9PL UK}
\address{$^2$ Argonne National Laboratory, Argonne, IL 60439 USA}

\begin{abstract}
{
The discovery that neutrinos have finite rest mass has led to renewed interest in neutrinoless double beta decay.   The development of 
    large-scale experiments to search for neutrinoless double beta decay 
has increased the probability of a 
credible observation of the process in the near future.  The reliability of calculations of the associated nuclear matrix elements is likely soon to become a critical issue. 
In this paper 
experimental techniques that access properties of the ground-state wave functions of double beta decay candidates, 
the occupancies of valence single-particle orbitals and pairing correlations, are summarized and the experimental data for candidate nuclei are reviewed. 
The results are discussed in relation to  questions concerning which aspects of nuclear structure may play an important role
in determining the nuclear matrix elements for neutrinoless double beta decay.  }
\end{abstract}

\submitto{\JPG}
\maketitle

\section{Introduction}							

With the discovery of neutrino oscillations, it has become evident that there are mass differences between the different eigenstates of neutrinos. This implies that neutrinos have a finite (though currently unknown) rest mass \cite{fkm99,Araki}, 
where the upper limit is currently 0.28 eV/c$^2$ \cite{th10}.   Majorana had suggested 75 years ago  \cite{maj} that the statistics for neutral, spin-1/2 particles might not be the same as those for charged ones, and allowing the possibility that neutrinos and antineutrinos could be one and the same.  As long as neutrinos had zero mass, chirality distinguished neutrinos from antineutrinos.  But with a finite rest mass established, the handedness of neutrinos is no longer an intrinsic property, making Majorana's hypothesis much more attractive, with the implication that the present concept of the conservation of lepton number comes into question.  

Double beta decay \cite{elvo} in which two neutrinos are emitted (2$\nu$2$\beta$) has been observed in a number of cases with lifetimes in the range of 10$^{19-25}$ years.  If the Majorana hypothesis is correct, it becomes feasible for two virtual neutrinos to annihilate and the resulting 0$\nu$2$\beta$ decay would proceed with the two electrons bearing the total decay energy \cite{elvo}.  The latter process is expected to proceed at a slower rates with lifetimes greater than 10$^{25}$ years, with the value depending on the nuclear matrix element and the electron neutrino rest mass, as well as the available energy.  The observation of this process may well be the only way to resolve issues on the basic nature of the neutrino and to provide information on the absolute value of its rest mass. 

To extract the mass, it will be crucial to have confidence in the reliability with which the nuclear matrix element can be calculated.  Most nuclear processes have been characterized by theoretical descriptions somewhat empirically, by looking at systematics and then developing a model
that can describe the systematics consistently. This provides a semi-empirical framework for further calculations.  For the neutrinoless double beta decay, we are unlikely to have this luxury since, even if it is observed, the number of cases are likely to be insufficient to establish systematics.  It is therefore desirable to obtain as much data as possible to characterize the relevant properties of the nuclei that are the most likely candidates for the observation of such decays.

The framework used for estimating the matrix elements has been largely that provided by the proton-neutron quasiparticle random phase approximation (QRPA)  \cite{elvo} which is successful in describing overall features of nuclear properties. 
QRPA depends on some rather drastic simplifying assumptions and is thus often not quantitatively accurate in describing the details of nuclear structure.  
On the other hand, large model spaces can easily be accommodated. Shell-model calculations are also used \cite{poves}. The shell model  can  generally describe most features of nuclear structure more quantitatively, but 
there are limits on the range of orbitals that can be included in the calculations. This restricted model space requires renormalisation of transition operators  and the closure approximation \cite{closure} is invoked for the virtual intermediate states in neutrinoless double beta decay.  Detailed aspects of  nuclear structure can also be 
described using so-called algebraic approaches, where the interacting boson approximation (IBA) has been found to be very useful. It is therefore not surprising that double beta decay matrix elements have also been calculated using the IBA model \cite{iac}, 
again utilizing the closure approximation.

From an experimentalistÕs perspective, the dominant question is to understand which readily measurable nuclear properties  probe aspects of the nuclear Hamiltonian that are  important for the nuclear matrix elements. These properties can then be measured and the results used to investigate the extent to which the theoretical models used for the matrix element calculations reproduce them.

In the past few years, we have carried out a series of experiments designed to test some of the nuclear properties that are likely to be relevant to the neutrinoless decay process.  In this article we review these experiments and the general methods. We also attempt to summarize the issues that remain in determining the relevance of these properties to the decay process.  This article is written from the perspective of experimental physicists trying to identify the relevant information.

\section{What Aspects Of Nuclear Structure Matter? }

The question of how easily observable features of  nuclei may be relevant to $0\nu2\beta$ decay is, somewhat surprisingly, not very well explored.  We give some qualitative arguments here -- but their validity needs to be investigated more rigorously.

\subsection{Relationship to other weak processes.}

The understanding of simple beta decay in nuclei is reasonably complete, especially for allowed transitions.  The phenomenal constancy of super-allowed transitions between mirror nuclei \cite{hardy} is one of the cornerstones of the field.  But decay rates to more complicated states, especially in nuclei close to regions of changing deformation, are more difficult to predict.

Matrix elements of  the 2$\nu$2$\beta$  decay process are calculated by allowing for  
the virtual excitation of states in the intermediate nucleus between the parent and daughter.  
Since the momentum provided by one of the beta decays is of the order of $\sim$1 MeV/c, only the lowest multipolarities (1$^+$ states) at the lowest excitation energies can contribute significantly.  If the energies and matrix elements of these intermediate states can be measured, for example by charge exchange reactions, a reasonable estimate can be made.  

For neutrinoless 0$\nu$2$\beta$ decay, however, there is a fundamental qualitative and quantitative difference.  Here the two neutrons decaying into  protons emit virtual neutrinos that must annihilate in a short distance within the nucleus \cite{vogel1} and such a distance implies high virtual momenta, up to several 100 MeV/c.  Consequently, the multipolarities involved in the intermediate virtual excitation \cite{suh1} can be quite high -- perhaps up to 8  -- and  excitation energies in the intermediate nucleus up to $\sim$100 MeV can play a role in the virtual process.   The sensitivity to the structure of the intermediate nucleus, which  is dominant for the 2$\nu$2$\beta$ process, is not very relevant here \cite{vogel2}.   Because of this difference, it does not seem reasonable to assume that a theory successfully describing the matrix elements for single or double beta decays with $real$ neutrinos represents any assurance that the same method can also reliably predict the neutrinoless mode.

\subsection{The relevance of the intermediate virtual excitations. }

Charge-exchange reactions such as ($^3$He,t) can excite particle-hole modes in the ``giant resonance'' region at excitation energies from 10 to 25 MeV.  The experimental measurements  of 1$^+$ excitations studied in this way have been crucial to calculations of the matrix elements for the  2$\nu$2$\beta$ mode that is dominated by the virtual excitation of a few 1$^+$ states at low excitation in the intermediate nucleus.    For the case of the 0$\nu$2$\beta$ process however, this is not the case because of the high virtual momenta available in the intermediate vertex.  This leads to many higher multipoles and excitation energies up to 50-100~MeV, encompassing all relevant particle-hole excitations.  For the neutrinoless mode, it is not at all clear that results from charge-exchange reactions are likely to be useful in constraining the calculations, since all possible modes are within easy reach.  The question of whether a closure approximation is appropriate has been investigated, and it has been found to be satisfied to better than 10$\%$ \cite{vogel1} at least in QRPA approaches.

To the extent that the closure approximation is valid, we can say that the details of the intermediate nucleus are not of major relevance in determining the matrix element for neutrinoless decay since the response function of nuclei to particle-hole excitations in the region relevant to this process has essentially all the possible modes accessible.

\subsection{Initial and final states. }

What nuclear properties can make a difference  for the 0$\nu$2$\beta$ decay?  The specific configurations of the valence nucleons that make up the ground states certainly $\it{have}$ to make a difference.  

All the candidates for such decays occur
between even nuclei with spin-parity 0$^+$.  The decay involves the destruction of a pair of neutrons and the creation of a proton pair.  But if there are additional changes to the configuration of nucleons, such as differences in the occupancy of the valence orbits greater than just one pair, the re-arrangement of many nucleons  could inhibit the decay, just as it does for other nuclear processes such as single $\beta$ decay or $\gamma$ decay, and illustrated schematically in Figure 1. Such effects are not within the present framework of QRPA calculations, but they do occur in real nuclei.
\begin{center}\includegraphics[width=140mm]{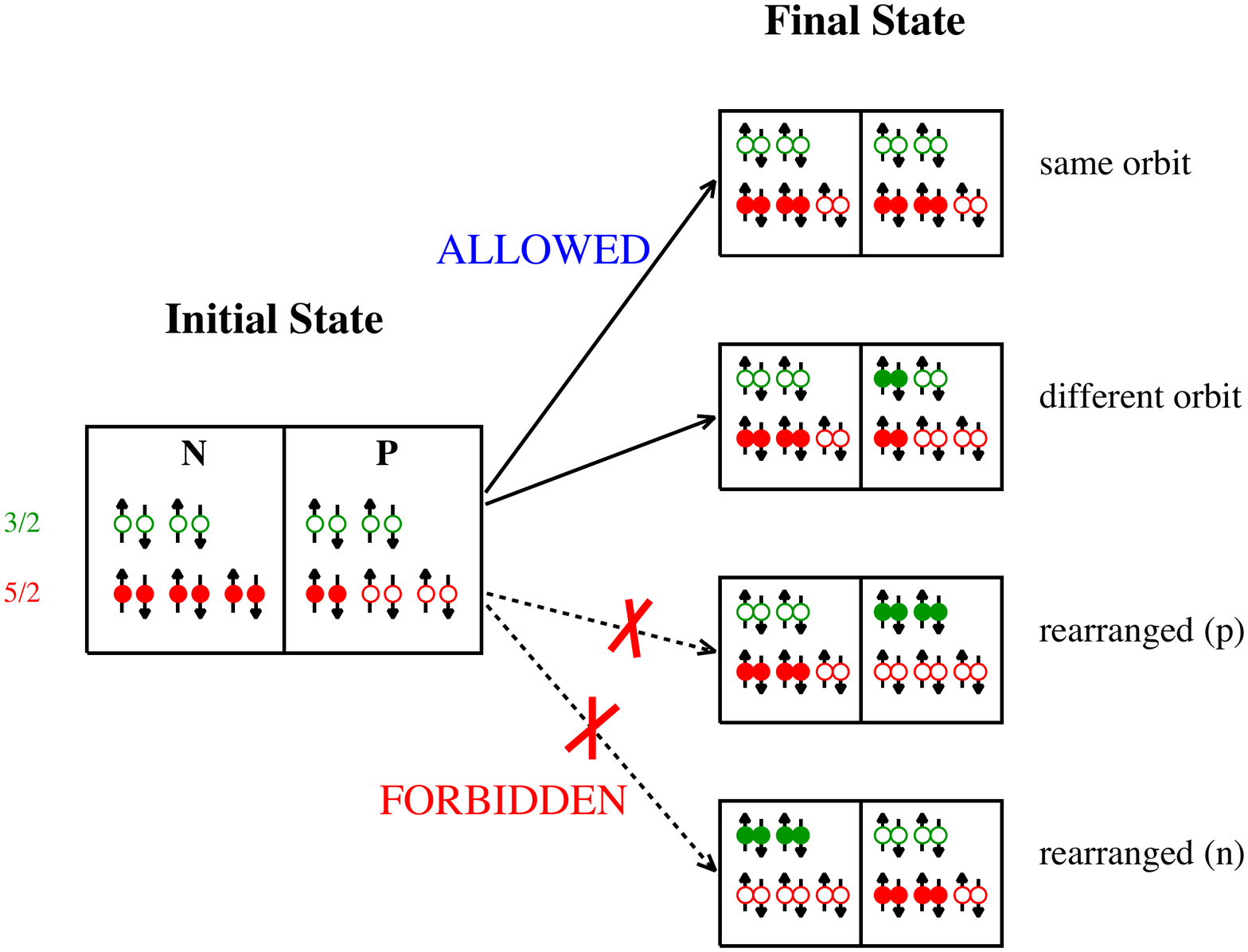}\end{center}

\noindent{\footnotesize {\bf Figure 1:} Caricature of how the rearrangement of nucleons would inhibit transitions.  The initial state is illustrated on the left, with just two valence orbits.  The two upper boxes on the right illustrate ``allowed'' transitions where a pair of neutrons decay into a protons pair in either the same or a different orbit.  The two lower boxes on the right schematically show configurations such as might arise from a rearrangement of valence orbits, changing deformations, or other effects.}
\vskip 0.5truecm

One example would be a change in deformation between the initial state and the final one.  This involves the rearrangement of many nucleons and is likely to inhibit the matrix element for the transition severely, perhaps by orders of magnitude, as it does in electromagnetic transitions in the Ôshape isomersÕ in heavy nuclei \cite{shape_isomer}.  
Inhibition based on the same mechanism can also occur in $\beta$ decay. For example, in the region of changing deformation around neutron number $N=90$, 
there are large order-of-magnitude variations in $ft$-values for transitions to various 0$^+$ states in the transitional nucleus  $^{152}$Sm, depending on whether the decay is from the $N=91$ 1$^+$ state of $^{152}$Pm, or from the $N=89$ 0$^-$ state of $^{152}$Eu, as is illustrated in Figure 2.

\begin{center}\includegraphics[width=140mm]{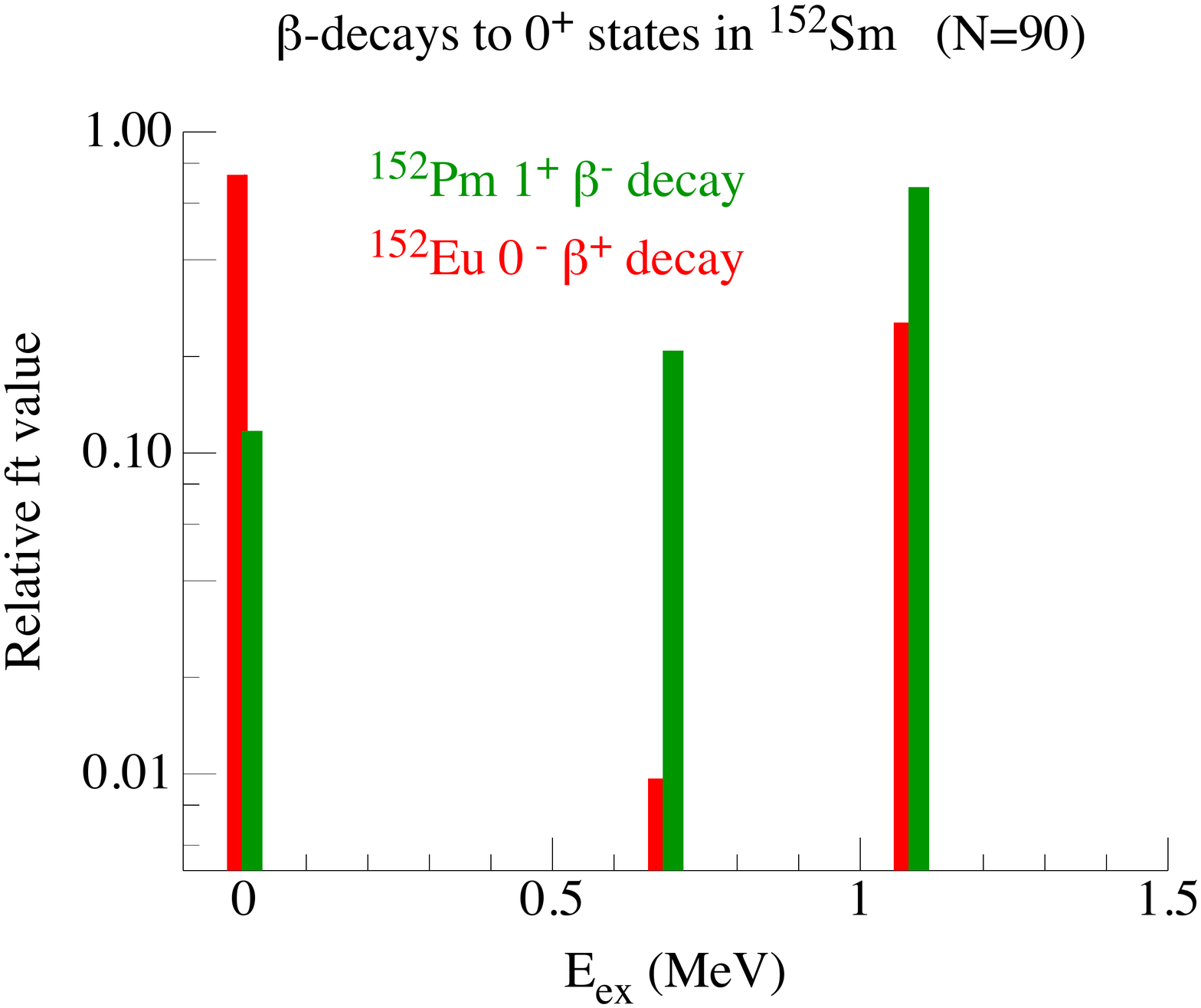}\end{center}

\noindent{\footnotesize {\bf Figure 2:} {The  ft-values for $\beta$-decay to the first three 0$^+$ states in the transitional $^{152}$Sm nucleus are drastically different, depending  on whether the decay is from the more deformed N=91 $^{152}$Eu, or from the more spherical N=89 $^{152}$Pm. For ease of display the ft-values for each decay were divided by the sum of the ft values to the three states.}}
\vskip 0.5truecm

In this paper we discuss two classes of measurements that we have carried out to characterize the initial and final states in 
the nuclei that are candidates for the observation of $0\nu2\beta$ decays: 
\begin{itemize}
\item the populations of various single-particle orbits of valence nucleons, with particular attention to the changes in occupancies, and
\item the extent to which correlations between the zero-coupled pairs of like nucleons, that lead to initial and final states consistent with the Bardeen-Cooper-Schrieffer (BCS) approximation, are confined to the ground states and are similar for the parent and daughter nuclei.
\end{itemize}

The details of how much specific differences affect the matrix elements are not yet clear, but some general considerations will apply.

\subsection{Valence Populations.}
The simplest way to consider differences between initial and final states is to look at the decomposition of these states in terms of populations of  single-particle orbits of valence nucleons.  The purpose is, first of all, to check that the changes did not imply some inhibition through Ôre-arrangementÕ effects and then to identify the specific orbits that are changing for both protons and neutrons. 

It has not yet been possible to gain a simple understanding of how specific orbits contribute to the process since, in the current QRPA formalism, their population can be controlled only indirectly, by modifying the energies of single-particle states for example.  What has been done \cite{vogel2} suggests that  the na\"{i}ve 
expectations are correct; for example, that the largest contributions to the matrix elements occur when the protons are created in the same orbits in which the neutrons decayed, for example $0g_{9/2}$ to  $0g_{9/2}$.

It is qualitatively obvious that the process will be inhibited when there are rearrangements in the configurations of nucleons other than the simple change implied by the decay, but there is considerable uncertainty in the quantitative estimates  of such inhibitions.

\subsection{Pair Correlations.}
Both modes of double beta decay involve the conversion of two neutrons into two protons, so 
it may perhaps be expected that correlations between nucleons may 
be relevant to the nuclear matrix elements for decay, as they are to pair-adding or pair-removing reactions. One important type of correlation is pairing, where the attractive nucleon-nucleon forces lead to the formation of $J^\pi=0^+$ nucleon pairs. At the simplest level, the displacement of the even-even and odd-odd mass parabolas due to the pairing energy is responsible for the very existence of double beta decay candidates. The attractive interactions localize the nucleon pairs spatially and distribute 
nucleons between single-particle orbitals close to the Fermi surface, which thus becomes smeared and orbitals acquire partial occupancy. 

If the 0$\nu\beta\beta$ matrix element is written as a sum over the angular momentum $J^\pi$ of products of neutron pair-annihilation and proton pair-creation operators, the part of the matrix element arising from nucleon pairs coupled to total spin $0^+$ or $J^\pi\neq0$ may be separated.  
Recent  calculations \cite{Engel88, Caurier, Escuderos} suggest that while the contributions via non-zero pairs, which have a long range,
are significant, they tend to have opposite sign to the contribution from the 0$^+$ pairs.  There are cancellation effects which diminish the long-range components leaving a short-range peak \cite{vogel2}, reminiscent of the wave functions of a pair of neutrons in $^3$H.

In shell-model treatments, pairing correlations are treated exactly, at least within the model space used in the calculation.
In QRPA methods, pairing correlations between {\it like} nucleons are treated separately from other effective interactions via a  transformation to a quasiparticle regime within the BCS   approximation \cite{BMP}. However, there are well-established circumstances in nuclei where the simple BCS approximation fails, as will be discussed in more detail below. How this may alter the calculation of matrix elements has not yet been explored.

\section{Single-Nucleon Transfer and Valence Populations }

In order to compare the differences in valence populations between initial and final states, the most suitable techniques are provided by single-nucleon transfer reactions.  In principle, knockout reactions provide an alternative, but these are not likely to be practical for the present purpose because of the energy resolution required. 
In nucleon-adding or nucleon-removing reactions,
the angular distributions are characteristic of the orbital angular momentum transfer and the cross sections are proportional to  spectroscopic overlaps between states.   These overlaps, or spectroscopic factors, were formally defined by Macfarlane and French \cite{macf} and  obey sum rules that relate their sums to the occupancies of single-particle orbits.  The summed spectroscopic factors for nucleon-adding reactions with a given quantum number reflect the vacancies in the corresponding orbit, while the sums for nucleon removing reflect the occupancy. 

\subsection {Normalizations~and~Absolute~Spectroscopic~Factors}

While there is some uncertainty in the absolute normalization of the reaction calculations, it has recently been shown that consistent results can be obtained by measuring both nucleon-adding and nucleon-removing reactions on the same target \cite{gese1, gese2, nisumrule}.   
The method consists of requiring a normalization such 
that for a given orbit characterised by total angular momentum $j$, the sum of the measured occupancy and vacancy on the same target  add up to the degeneracy of the orbit $2j+1$.  These measurements were carried out with special attention to minimizing errors in cross sections in order to reduce systematic uncertainties.  The neutron-transfer reactions were chosen to optimize momentum matching, using  ($\alpha,^3$He) reactions for higher $\ell$ values and (d,p) reactions for lower values.  The analyses were  carried out within the Distorted Wave Born Approximation (DWBA) with the distortions taken from global optical potentials.  This procedure leads to an independent value of the normalization for each target nucleus, and the consistency between four independent determinations is illustrated in Figure 3 for the stable Ni isotopes.  The four values were then averaged, so that the same normalization was used for all the neutron-transfer measurements on Ni.
This same normalization also gives the neutron occupancies of the valence orbits, constituting an independent check of the consistency.
These results were not very sensitive to the choice of potential parameters in the reaction modelling as long as a consistent procedure was used; the uncertainties arising from the 
choice of potentials were at a level of a few percent in the summed spectroscopic factors and under 10$\%$ for the individual spectroscopic factors. This normalization procedure removes a large part of the uncertainty in using transfer reactions to obtain occupancies; it relies on reaction theories primarily for the rather small changes in comparing reactions with different Q-values.

\begin{center}\includegraphics[width=140mm]{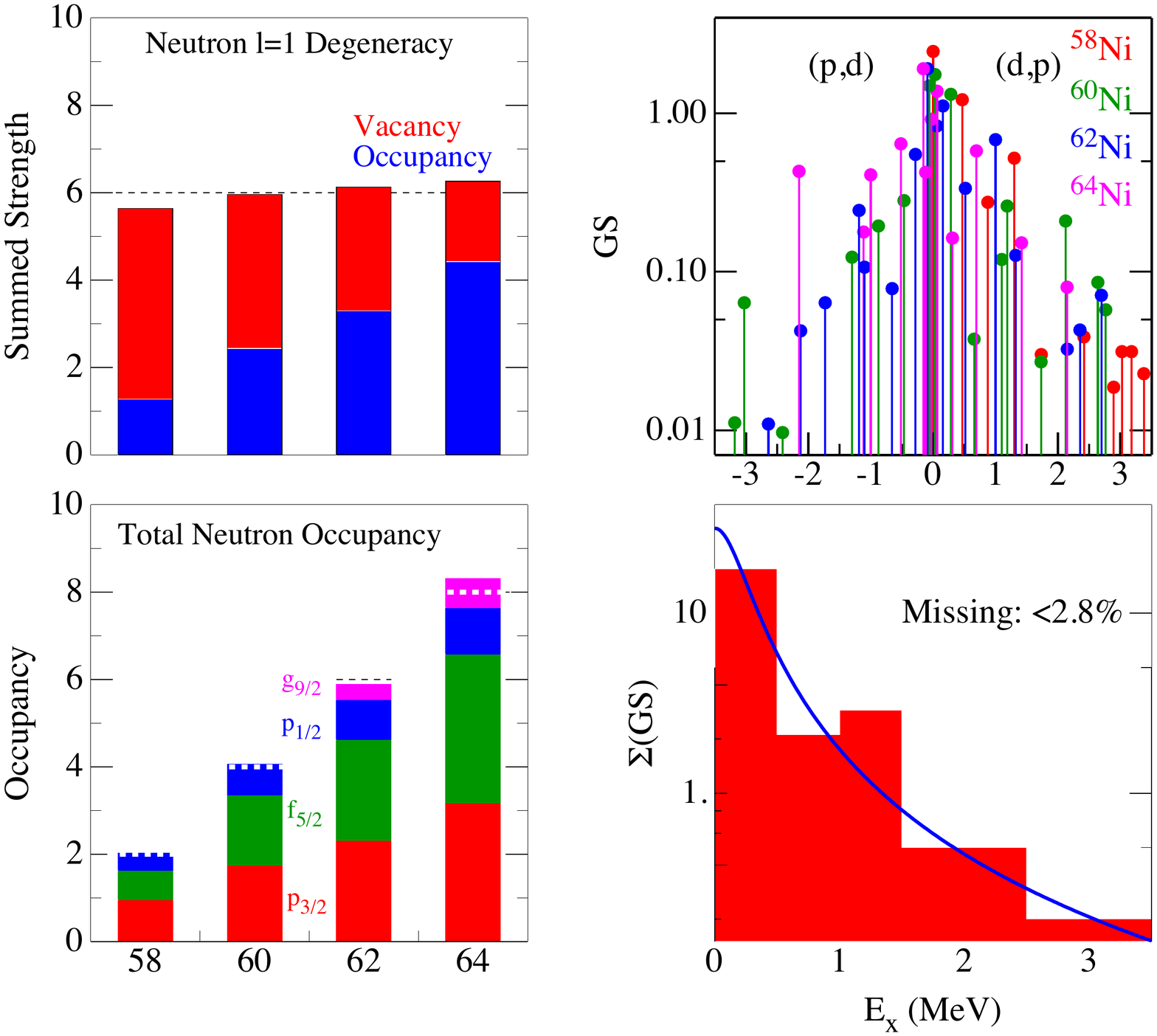}\end{center}
\noindent{\footnotesize {\bf Figure 3:} {Results from single-neutron transfer reactions on four Ni isotopes.  The box on
the top left illustrates the consistency in the degeneracy, once the
normalized  $\ell$=1 spectroscopic factors are summed for both occupancies
and vacancies.  The lower left shows the changing neutron occupancies in
the valence orbits derived from the transfer data, and their consistency
with expectations, with the dashed line indicating the expected levels of
valence-neutron occupancies.  On the top right the fragmentation of the
$\ell$=1 strength is shown, as a function of excitation energy, using
negative energies for hole states from neutron-removal.  The bottom right
shows a histogram with the binned strength fitted to a Lorentzian, to
indicate that not much of the total strength is likely to be missed.} }
\vskip 0.5truecm

It has been previously noted, particularly for (e,e$^\prime$p) reactions, that $\it{absolute}$ spectroscopic factors fall short of the single-particle value, and this has been understood in terms of the limitations of the mean-field approximation arising from high-momentum short-range correlations between nucleons \cite{lap, pan}.  The normalizations obtained from the transfer data by the above 
procedure 
are similar to the values expected on the basis of the quenching of low-lying single-particle strength observed in (e,e$^\prime$p) experiments.  
The value of this renormalization is apparently a uniform property of nuclear matter, and appears to be independent of nucleus, at least to the extent that the limited range of target nuclei allows us to make such tests.

The magnitude of the renormalisation adopted for the occupancy measurement was the average of the values determined for individual targets in a given vicinity of nuclei. The variation between the values was a few $\%$, as  shown in Figure 3.  The consistency of the normalizations, and the fact that they lead to sensible values of the occupancies to a few tenths of a nucleon, indicates the level of validity in such a procedure for determining changes in valence occupancies.  The normalizations depend slightly on the optical-model parameters used, but the normalized spectroscopic factors are essentially the same for different ``reasonable'' parameters.

A great deal of information is available from transfer reaction studies performed since the 1960s, which established the foundations of our current understanding of nuclei and the single-particle skeleton underlying nuclear structure. Unfortunately, cross sections were often not published; in many cases only graphs of angular distributions and tables of spectroscopic factors  are available.  The analyses were carried out with a variety of parameters, computer codes, and approximations.  Thus the precision and consistency needed for quantitative comparisons of data on different nuclei is not readily available from the published literature, even where the relevant measurements were carried out with great care.  

For the purposes of meaningful comparisons of valence occupancies between initial and final states in double beta decay, the older experiments could only be used as a guide for identifying $\ell$ values and to indicate which transitions were strong and which were weak.   An example 
of the type of information that has been available is shown in Figure 4, together with a point indicating the accuracy in absolute cross section obtained in \cite{gese1}.

\begin{center}\begin{center}\includegraphics[width=110mm]{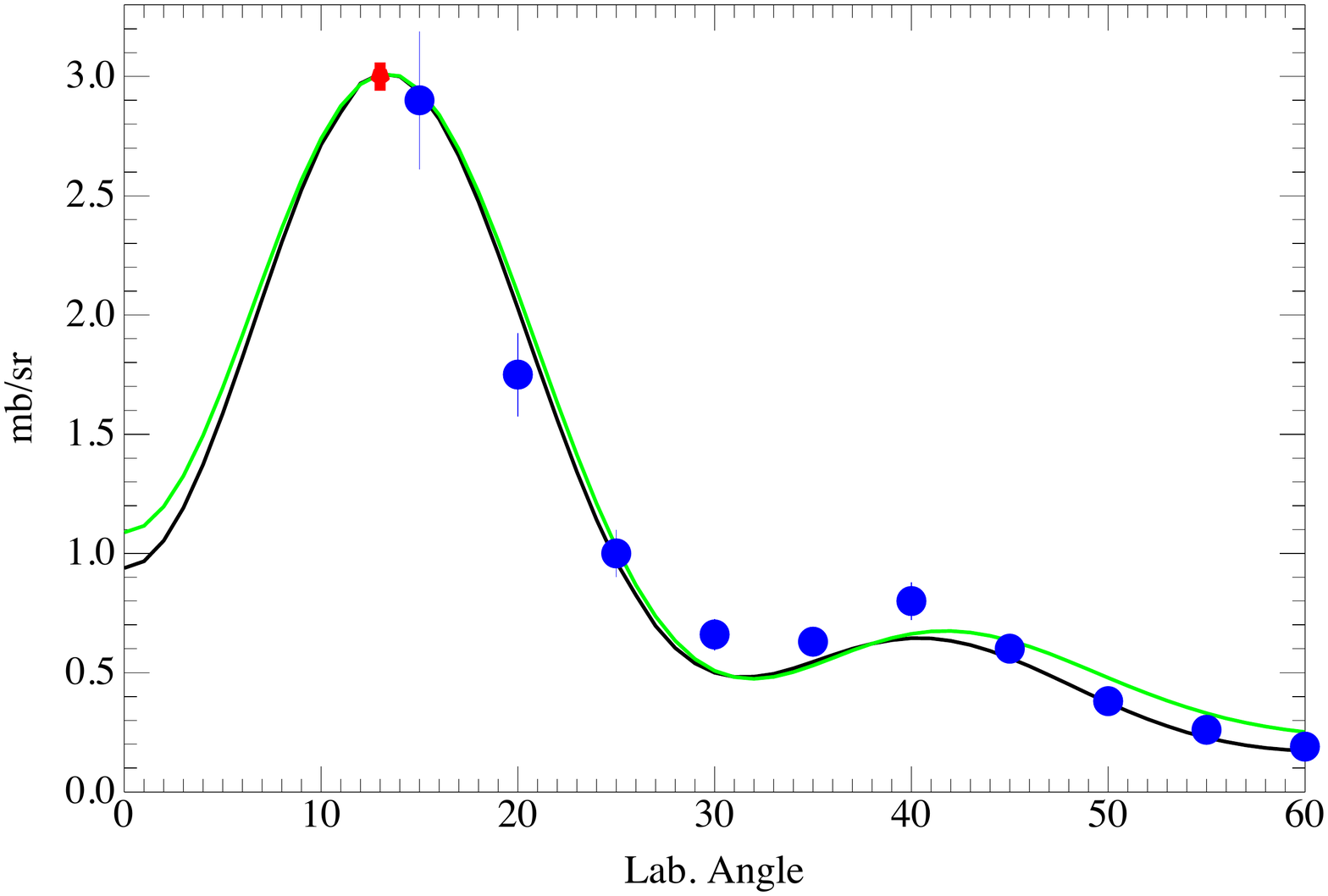}\end{center}\end{center}

\noindent{\footnotesize{\bf Figure 4:} {  
Data taken in an earlier measurement \cite{yoh}
for a transition in $^{76}$Ge(d,p)$^{77}$Ge to the 0.159-MeV excited state. The blue points with error bars are values obtained from the figure in this reference and their stated absolute errors.  The green and black lines represent two DWBA calculations with different global distorting parameters.  The red bar illustrates the fractional accuracy in the measurements that were carried out to help constrain the $0\nu2\beta$ matrix element \cite{gese1}.   } }
\vskip 0.5truecm

\subsection{Summary of what has been done, what is under way, and what remains.}

In this section we will review what has been done in determining changes in valence populations of nuclei that are candidates for the observation of $0\nu2\beta$ decay.  The major experiments searching 
 for this exotic decay mode that are currently under way utilize $^{76}$Ge, $^{100}$Mo, $^{130}$Te, $^{136}$Xe, and $^{150}$Nd.  Other candidate decays under consideration or being tested for suitability are $^{48}$Ca, $^{82}$Se, $^{96}$Zr, $^{116}$Cd, etc.  We have completed the measurements on determining the valence-nucleon populations in $^{76}$Ge, and are well underway for $^{100}$Mo and $^{130}$Te systems, with measurements associated with $^{136}$Xe and $^{150}$Nd planned for the near future.  In this section we discuss the status of the experiments undertaken so far. 

\subsubsection{The valence orbit occupancies in $^{76}$Ge and $^{76}$Se.}
$\newline$
The study of this system has been completed and was published in Refs. \cite{gese1, gese2}.   Some of the general considerations are discussed in more detail for this system in order to illustrate the issues involved with all the studies.  The germanium and selenium isotopes had  been studied with (d,p) reactions several decades ago, such that $\ell$ values and approximate spectroscopic factors  were known \cite{ensdf}.  Many of the levels that are populated in the transfer reactions had also been studied by $\gamma$-ray spectroscopy and thus spins and parities were confirmed.  The active neutron orbits are $1p_{3/2}$, $0f_{5/2}$, $1p_{1/2}$, and $0g_{9/2}$, with 44 neutrons in $^{76}$Ge. 

For a given $\ell$ value, the calculated angular distributions of (d,p) reactions predict the peak angles to a reasonable accuracy and the cross sections in Refs. \cite{gese1} were measured with good statistics at these angles.   The same procedure was followed for the (p,d) reaction, where the bombarding energy was chosen to yield proton and deuteron energies as close to those in the (d,p) reaction as possible.  To obtain spectroscopic factors we rely on the DWBA model primarily to give the energy dependencies.  Data were obtained for four target nuclei to provide some redundancy in the measurements: $^{74,76}$Ge and $^{76,78}$Se, with 32 and 34 protons for Ge and Se and 42 and 44 neutrons for the two isotopes chosen.  

A test similar to that shown in Figure 3 was carried out to give confidence in having covered all strengths.  Requiring the sum of vacancies and occupancies to be $2j+1$ yielded very similar normalizations: 0.53$\pm$.02 for $\ell$=1 and 0.54$\pm$.06 for $\ell$=3.  In a chronologically later experiment on four nearby Ni isotopes that was already mentioned \cite{nisumrule}, comparable values were obtained: 0.57$\pm$.02 and 0.52$\pm$.03 respectively. This consistency is further evidence indicating that the procedure followed is valid.  The results for the valence occupancies in the Ge and Se nuclei are shown in Figure 5.

\begin{center}\includegraphics[width=120mm]{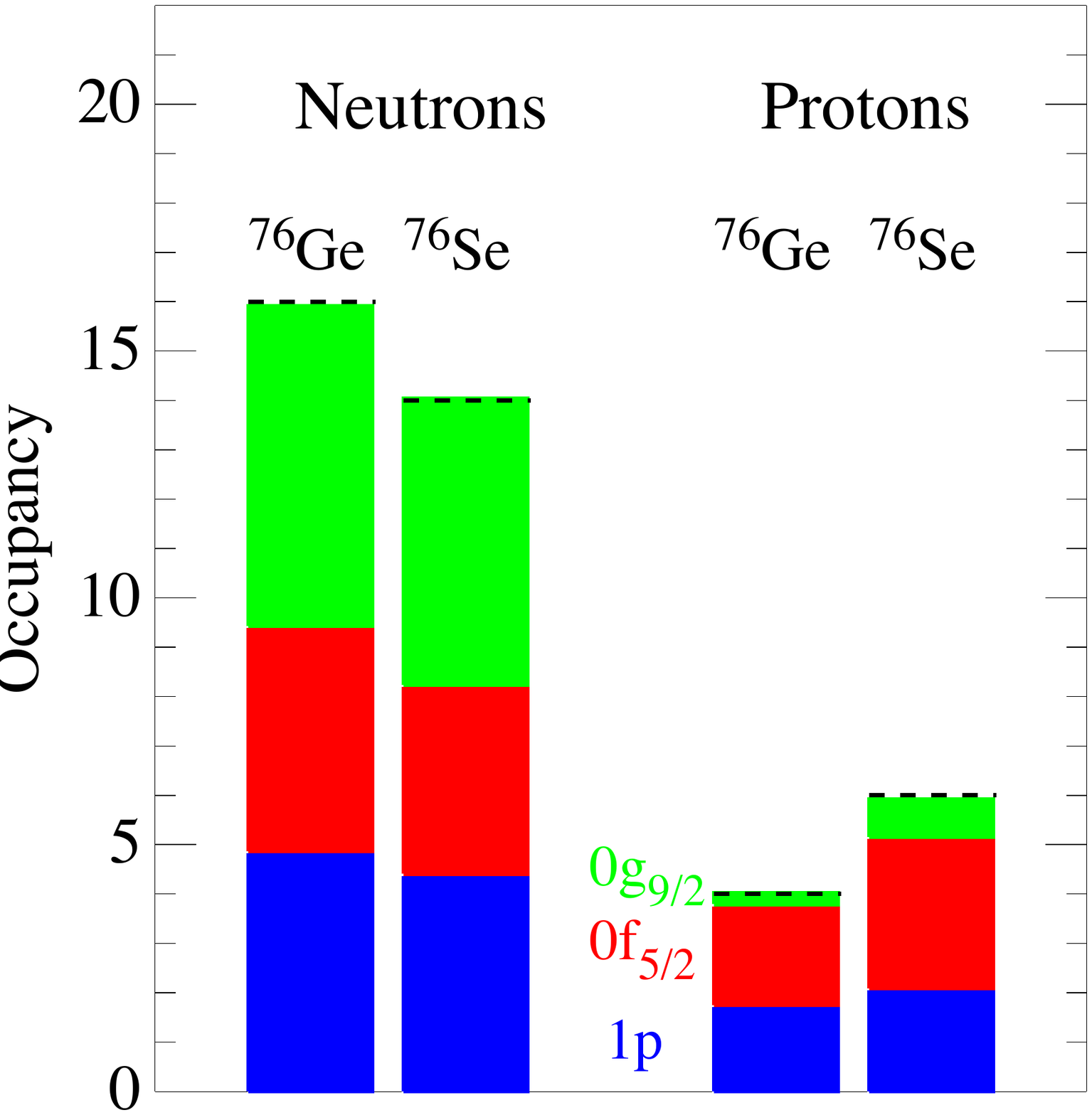}\end{center}

\noindent{\footnotesize {\bf Figure 5} {Occupancies for neutron and proton orbitals  in $^{76}$Ge and $^{76}$Se measured in Ref.s \cite{gese1, gese2}. Not all the spins for the $\ell$=1 transitions were known unambiguously and therefore 
only the total $\ell$=1 strength is shown.}}
\vskip 0.5truecm 

The proton spectroscopic factors were obtained in a similar fashion, using ($^3$He,d) and (d,$^3$He) reactions \cite{gese2}.  However, Ge and Se have 32 and 34 protons respectively, just four and six protons beyond $Z=28$, and well below $Z=50$, in the same major shell as the neutrons.  Thus the proton-removing (d,$^3$He) reaction is critical for the experiment. Since the Z=50 closed shell is far away, the centroid of the transition strength for proton addition will be at higher excitation energy (over 3 MeV for $9/2^+$) and tend to be fragmented into many small components.  In contrast, for the neutron reactions with 42 or 44 neutrons, these nuclei are about midway between shells and the neutron-transfer centroids are well below 2 MeV excitation.   The normalization procedure for proton transfer therefore had to be somewhat different from that followed for neutrons.
   
The (d,$^3$He) measurements were done at the RCNP in Osaka with the Grand Raiden spectrograph \cite{grandraiden}.  At this facility the lowest practical energy was 80 MeV for the deuterons, which still provides  reasonable momentum matching for the orbits of interest. For the (d,$^3$He) reaction, the momentum matched well for $\ell$ about 2.5, and still reasonable for 1 and 4, thus suitable for the  transitions to $p$, $f$ and $g$ orbitals of interest. 
Assuming a single normalization for the three $\ell$ transfers, one can normalize the DWBA calculations by requiring that the occupancies be equal to 4.0 for germanium and 6.0 for selenium, providing a fourfold redundancy in the normalization.  This yielded proton occupancies of 3.8, 4.0, 6.1 and 6.2 for the four targets, $^{74,76}$Ge and $^{76,78}$Se, indicating consistency at the level of around 0.2 nucleons.  

A comparison between the differences in orbital occupancies for neutrons and protons from our measurements are shown in Figure 6.  The QRPA calculations that were done before \cite{qrpa1} and after \cite{qrpa2} publication of the experimental results are also shown in the figure.  It is clear that the prior calculations did not fit the measurements, but adjusting the assumed single-particle energies in the QRPA considerably improved the agreement and the calculated decay rate changed by about a factor of two.

\begin{center}\includegraphics[width=120mm]{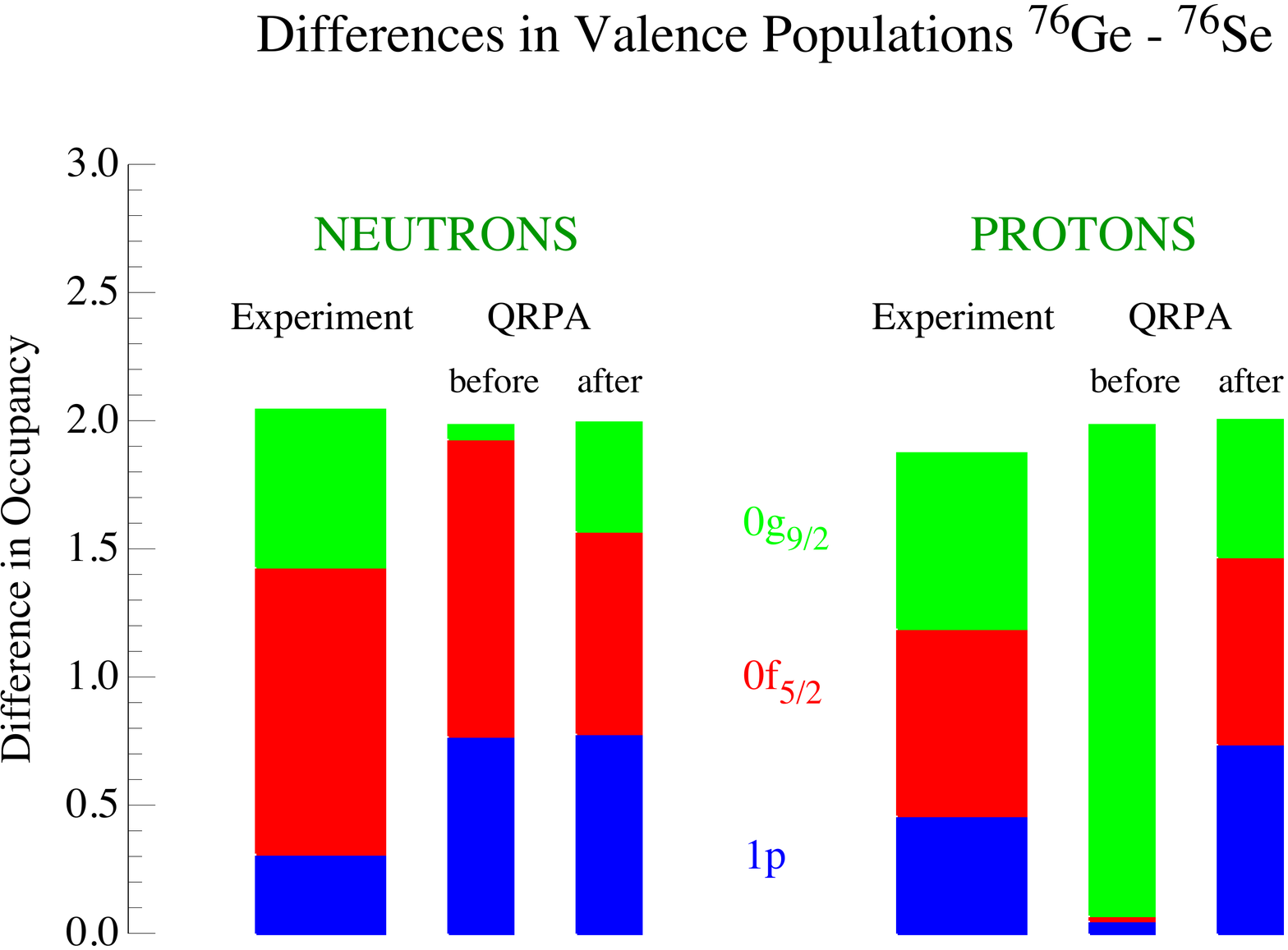}\end{center}

\noindent{\footnotesize {\bf Figure 6:} {Changes in the neutron and proton populations of valence orbits from the measurements in Ref.s \cite{gese1,gese2}, along with those from QRPA calculations from before and after the transfer-reaction measurements were published.  The changes are shown as positive for both protons and neutrons, even though the neutron number decreases.}}
\vskip 0.5truecm 

\subsubsection{The valence orbit occupancies in $^{130}$Te and $^{130}$Xe.}
$\newline$
This measurement has been started, but the data analysis is not yet final \cite{texe130}.  The neutron occupancies were determined for $^{128,130}$Te with similar techniques to those described in the previous section.  Since these nuclei are close to the $N=82$ neutron shell, with 4 and 6 vacancies, the critical measurements are the determinations of the neutron vacancies by neutron-adding reactions.  The orbits that are important are 1d$_{3/2}$, 1d$_{5/2}$, 0h$_{11/2}$, and possibly 0g$_{7/2}$.  These reactions had been previously studied and the strengths seemed to be concentrated in the low-lying states.  Because some of the $\ell$ values were high, the ($\alpha,^3$He) reaction was important to achieve the required momentum matching.  
For the inverse ($^3$He,$\alpha$) reaction, the $\ell$=5 strength was found to be in a single state, but  the $\ell$=2 strength was found to be  fragmented and apparently spread to high excitation energy, so that it could not be determined reliably.  Using a normalization deduced only for $\ell$=5 and the requirement that the total neutron vacancies add up to 4 or 6, consistent results were obtained.   No evidence for vacancies in the g$_{7/2}$ orbit 
was seen.

For the measurement on  xenon isotopes, in order to get the thin targets required to provide adequate resolution and to avoid the windows associated with a gas-cell target, a frozen layer of xenon was formed by spraying gas onto a thin diamond foil held at a low temperature \cite{bloxham}.  Although this limited the measurement to low beam currents, reasonable data were obtained, albeit with somewhat lower resolution than in the Te measurement, 
which used conventional targetry.  The summed strengths in both cases were consistent with the same normalization.  The results on the difference in neutron occupancies are shown in Figure~6.  

Experiments to determine the proton occupancies and changes in proton occupancies are in the planning stage.  This is particularly interesting in view of the apparent pair vibration that was observed in ($^3$He,n) studies that will be discussed in the Section 4.

\begin{center}\includegraphics[width=120 mm]{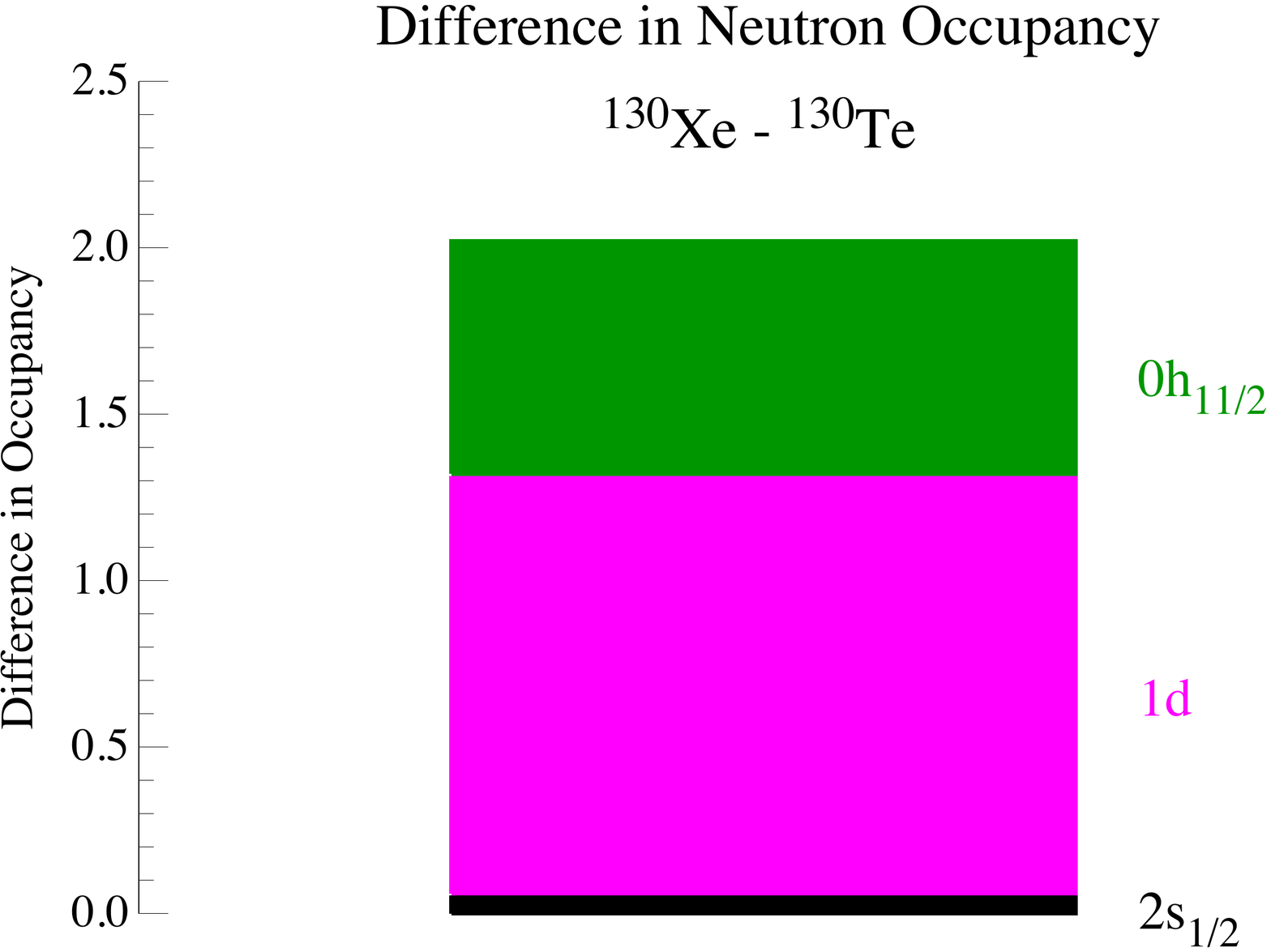}\end{center}
{\noindent\footnotesize{\bf Figure 7:} { Changes in the measured neutron populations of valence orbits between $^{130}$Te and $^{130}$Xe. The $\ell$=2 strength is probably mostly in the d$_{3/2}$ orbit, but not all the spins are known, so this is labelled 1d.}}
\vskip 0.5truecm 

\subsubsection{The valence orbit occupancies in $^{100}$Mo and $^{100}$Ru.}
$\newline$
High resolution measurements of  proton-adding and neutron-removing reactions have been carried out using the Q3D spectrometer at the Maier-Leibnitz Laboratory in Munich~\cite{moru100}.  The reactions were on targets corresponding to the double beta decay parent and daughter nuclei, and also on $^{102}$Ru, which served as a consistency check and to help estimate uncertainties better. The (p,d) and ($^3$He,d) reactions were used to obtain good momentum matching for low $\ell$ transfers, and ($^3$He, $\alpha)$ and ($\alpha$,t) reactions used to extract information for higher $\ell$ values. 

To normalize the spectroscopic factors, the total vacancies for valence protons in Mo and Ru were required to be equal to 8.0 and 6.0 respectively, and the total occupancy for valence neutrons beyond N=50 to 8.0 for $^{100}$Mo and $^{102}$Ru, and 6.0 for $^{100}$Ru. The analysis is almost complete and provisional results for the  occupancies have been presented  in Ref. \cite{moru100}.
 
\section{Pairing Correlations and Two-Nucleon Transfer Studies}

\subsection{Pairing, Two-Nucleon Transfer, and BCS.}
Since the most important pair correlations in nuclei are those manifest in the short-range spatial correlation of 0$^+$ pairs of identical nucleons, a single-step reaction transferring two such nucleons is a specific probe of pairing. Indeed, light-ion induced two-nucleon transfer reactions have been used for many years to experimentally access pairing effects. Most of the existing data is for the two-neutron-removing (p,t) reaction at bombarding energies of several 
tens of MeV.  Considerable information was also obtained for the neutron-pair-adding (t,p) reaction in the period from $\sim$1970-90, but this has diminished in recent years due to the increased regulations governing the use of radioactive triton beams. The analogous light-ion proton-pair-adding  reaction ($^3$He,n) is comparatively less well studied; the constraints of time-of-flight neutron detection limit the energy resolution, and while conditions were optimized in several dedicated facilities, such setups are not commonly found today.  Not all systems had been studied using ($^3$He,n) as carefully as they were for (p,t) and (t,p) reactions.

The usefulness of two-nucleon transfer in nuclear spectroscopy was realized by Yoshida \cite{Yoshida}, who first analyzed 
spectroscopic amplitudes for two-neutron transfer reactions in a Born approximation. In both $^3$H or $^3$He, there is a pair of  
identical s-wave nucleons coupled to 0$^+$.  
The cross section for pair transfer is enhanced when it is to, or from, a state in which BCS correlations result in a coherent mixture of pairs. A 0$^+$ pair of correlated nucleons is well localized and thus has good overlap with the pair in the mass-3 nucleus.  
The resulting collective enhancement in the reaction amplitude represents the participation of pairs in the various valence orbitals, adding with a common phase.   
Further, the binding energy of mass-3 nuclei is such that the momentum change needed to transfer an $\ell$=0 pair is favorable.  Simple quantitative estimates of the enhanced ground-state yield, compared to a  two-quasiparticle excited state, yield ${\sigma_{\rm gs\rightarrow gs} / \sigma_{\rm gs\rightarrow 2qp}}\sim {A/ 4}$ \cite{Yoshida, Brink/Broglia}.  For a medium mass nucleus, one may estimate that the cross section between ground states described by two fully-paired BCS wave functions may be enhanced by a factor of around 20--30 compared to the population of other non-collective 0$^+$ excited states.  

The gross features of a two-nucleon transfer spectrum should therefore be the strong population of the ground state of the final system, with  excited 0$^+$ states populated less than a few $\%$ of the ground state, assuming that the ground states of target and residual nucleus are well described by BCS wave functions.

\subsection{Departures from Simple BCS Behavior, Changing Shapes and Pairing Vibrations.}

The appearance of significant population of excited 0$^+$ states would indicate a departure from this simple picture and there are two common situations when this can occur: where there are 
gaps in the single-particle structure that are larger than the pairing interaction responsible for the correlations, and where there is a change in the shape of the nuclear ground-state. 

Gaps in the single-particle structure close to the Fermi surface gives rise to the phenomenon called pairing vibrations, the name arising from their discussion in terms of the collectivity associated with fields that promote a pair from one correlated state to another~\cite{Bes66}. Microscopically, the presence of a gap with an energy larger than that associated with the pairing interaction effectively isolates the sets of levels above it from being included in the correlated mixture that forms the ground state.
Depending on the details of structure, these higher single-particle states may then produce another excited, pair-correlated 0$^+$ state at higher excitation energy, with significant cross section for pair transfer. The effect of the gap is therefore to share the observed BCS pair-transfer strength between the ground and the excited 0$^+$ state. 

Near to significant closed shells such as in $^{208}$Pb, the robust energy gap gives rise to well-developed pair-vibrational behaviour and the
(p,t) and (t,p) reactions populate a near-harmonic set of pairing-vibrational states in neighbouring even-even Pb isotopes~\cite{lead_pairing_vibrations}. Sub-shell gaps in spherical or deformed single-particle energy levels larger than the pairing interaction, and/or variations in the strength of the pairing interaction among orbits, can also be responsible for pairing-vibration effects. The presence of such pairing-vibration departures from a simple BCS behavior 
indicates that the quantitative validity of the simple BCS approximation in the description of the ground-state wave functions is 
altered. This may well have important implications for the calculation of matrix elements for double beta decay, perhaps at a level comparable to or larger than the splitting of strength seen in two-nucleon transfer.

In regions of nuclei that span a change in ground-state shape, strong population of excited 0$^+$ states can occur where the overlap between the initial target wave function with a final excited-state wave function is large compared to that with the final ground-state wave function. Such effects arise in transitional regions, where 
the ground-state shape of the residual nucleus is different from that of the target. 
In such cases, there is often an excited 
state in the final system whose shape is more similar to 
that of the initial nucleus than the ground state, and thus strong pair transfer is facilitated to the excited state. 
The classic example of this phenomenon is at N=90, and actually involves one of the candidates for double beta-decay studies.  It will be discussed in more detail below;  
the ill-defined and changing shape will complicate calculations of matrix elements.

For measurements relevant to double beta decay, two-nucleon transfer experiments focus on the population of $0^+$ states, starting either with target nuclei that correspond to a double beta-decay parent or to a daughter. Excited $0^+$ final states stand out unambiguously because their angular distributions are uniquely forward peaked and the cross sections are the measure of spectroscopic overlap.  If the final states are more than a few 
percent of the ground-state transition,  the pair-removal process in double beta decay may also be fragmented and not well described by a theory that 
fails to take the changing shape into account explicitly.

\subsection{Experimental Information on Pair Transfer in Candidate Nuclei.}

The available information on pair transfer for double beta decay candidates is summarised here.

\subsubsection{$^{76}$Ge--$^{76}$Se}
$\newline$The nuclear structure of even Ge and Se isotopes close to stability is rather well studied and evidence for excited $0^+$ states has been established. In some of the lighter isotopes, two-neutron transfer reactions have shown significant strength populating some of these states and this has been interpreted in terms of shape coexistence effects.  There are no such effects in the A=76 system. 

Recent work \cite{Freeman} was aimed at studies in the vicinity of $^{76}$Ge for double beta decay using the (p,t) reaction on targets of $^{74,76}$Ge and $^{76,78}$Se.  The experiment was focused on measuring accurate cross sections at forward angles 
near the peak of the $\ell=0$ angular distribution. Typical spectra are shown in Figure 8. 
For the  targets that are relevant to the double beta decay, $^{76}$Ge and its daughter $^{76}$Se, there is no population of excited $0^+$ states beyond a few percent of the yield of the ground-state transitions. The experimental cross sections for the two nuclei, measured for the ground-state transitions,
 are remarkably constant within the 5\% experimental uncertainty. These features suggest that 
 for the neutrons in  $^{76}$Ge and $^{76}$Se 
 the quantitative nature of the pair correlations does not change appreciably between mother and daughter, and that the BCS approximation is reasonable.

\begin{center}\includegraphics[width=120mm]{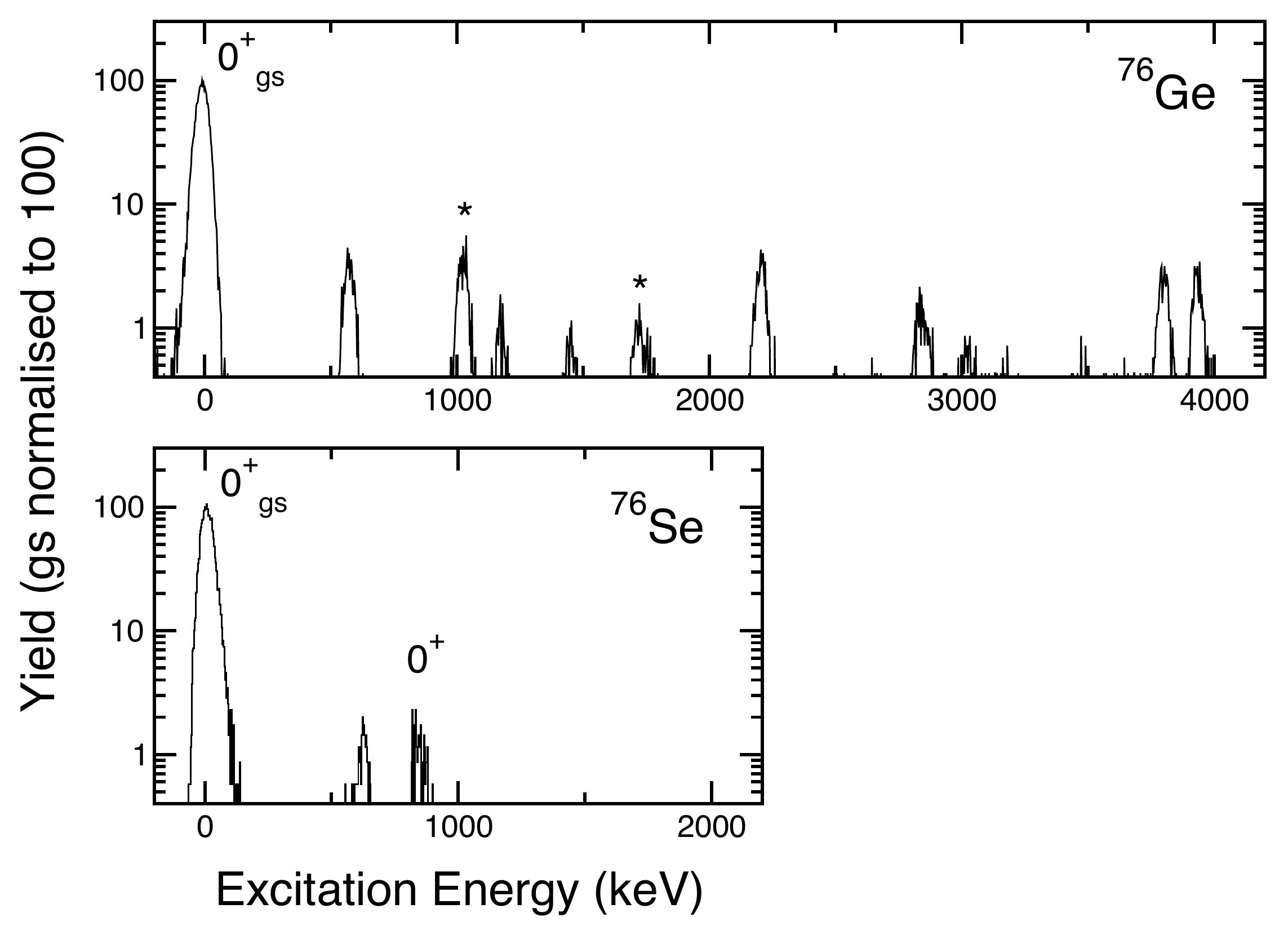}\end{center}
\noindent{\footnotesize {\bf Figure 8:} { 
Excitation energy spectra for the $^{76}$Ge(p,t) and   $^{76}$Se(p,t) reactions from experiments described in Ref. \cite{Freeman}. States with angular distributions consistent with 0$^+$ states are labelled. Peaks labelled with an asterisk are target contaminants.  There are no excited 0$^+$ states in either spectrum at a level of more than a few percent of the ground-state transition. }}
\vskip 0.5truecm 

There have been no studies of proton-pair transfer on these nuclei.  However, experiments are currently underway at Notre Dame University to study the ($^3$He,n) reaction on targets of $^{74,76}$Ge \cite{Jim}.

\subsubsection{$^{130}$Te--$^{130}$Xe.}
$\newline$ In the $^{130}$Te--$^{130}$Xe double beta decay system, there is some prior information on pair-transfer reactions. The most recent neutron-pair transfer work \cite{bloxham_te} focused on the $^{128,130}$Te(p,t) reactions in the forward angle region that is most sensitive to the population of 0$^+$ states. 
The measurement shows no strong population of excited 0$^+$ states, they were all  $<$4\% of the yield of the ground state.  Measurements were also made of the 
$^{132,130}$Xe(p,t)$^{130,128}$Xe reactions with a frozen xenon target and, while the analysis is not yet complete, it seems qualitatively consistent with what was seen with tellurium targets.

However, the situation for proton pairing in tellurium nuclei is very different. The ($^3$He,n)  proton-pair-adding reaction populates at least one excited 0$^+$ state on targets of $^{122-130}$Te with significant strength, $\sim$40\% of that of the ground-state transition \cite{alford1}. This striking departure from the BCS expectations is an example of  proton pairing vibrations.  The major shell between Z=50 and  82 has an additional gap, apparently separating the first 14 nucleons in the  $g_{7/2}$ and $d_{5/2}$  orbitals from the $h_{11/2}$,  $s_{1/2}$ and $d_{3/2}$ with 18 nucleons, creating a subshell at $Z=64$.  This gap is apparently sufficiently large, such that the orbitals above $Z=64$ cannot participate in the correlated ground state and thus do not contribute to the coherent BCS state.  This should become more apparent in the planned proton occupancy measurements mentioned in Section 3.2.2.  Such fragmentation of proton BCS strength in these systems  raises substantial questions regarding the calculations of matrix elements for double beta decay with QRPA, where a full BCS state is assumed as a starting point.  Shell-model calculations have also been carried out for the $A=130$ system \cite{poves}, but it is not clear whether these successfully reproduce the observed tellurium proton pair vibrations.

\subsubsection{$^{100}$Mo--$^{100}$Ru}
$\newline$There is considerable evidence  that there is a transition from spherical to deformed shapes for nuclei in the $A\sim 100$ region near $N=60$ with well-developed ground-state rotational bands in $A \ge 100$ in zirconium isotopes and $A \ge 102$ in molybdenum. The effects of this are clearly evident in experimental data on (p,t) and (t,p) reactions in the transitional region around the $^{100}$Mo double beta decay nucleus and a recent measurement has focused on a quantitative comparison of the relevant nuclei  \cite{jeff}. 
This measurement, using (p,t) reactions on targets of $^{98,100}$Mo and $^{100,102}$Ru,  found that $\ge$~95\% of the neutron pair transfer strength is contained in the ground-state transition, except for the reaction leading to $^{98}$Mo. In this case it was observed, in common with previous studies, that a state at 735~keV was populated with $\sim$20\% of the ground-state transition strength. The transitional nature of the $^{100}$Mo double beta decay parent nucleus is likely to present challenges for calculations of the matrix elements.

Some ($^3$He,n) data exist for stable $A\sim100$ nuclei \cite{fielding}. The data on targets of $^{100}$Mo and $^{102}$Ru indicated no evidence of excited 0$^+$ states, confirming the simple expectation for the proton configurations in the associated ground states.

\subsubsection{$^{150}$Nd--$^{150}$Sm}
$\newline$
Data for Sm(t,p) reactions \cite{bjerregaard_Sm} showed the first evidence for significant population of excited $0^+$ states in a region characterised by a transition between two different ground-state shapes using pair transfer reactions.  This is evident in subsequent experiments  both for neutron-pair adding and removing reactions on samarium isotopes \cite{ensdf}, with similar phenomena seen in neodymium isotopes \cite{chapman, yagi}.  The data can be interpreted in terms of a preference in the reaction to populate a final state with a deformation similar to that of the initial state.  
The effect in pair transfer reactions is most pronounced between nuclei with 88 and 90 neutrons; this is identical the change in neutron number involved in double beta decay.

\begin{center}\includegraphics[width=100mm]{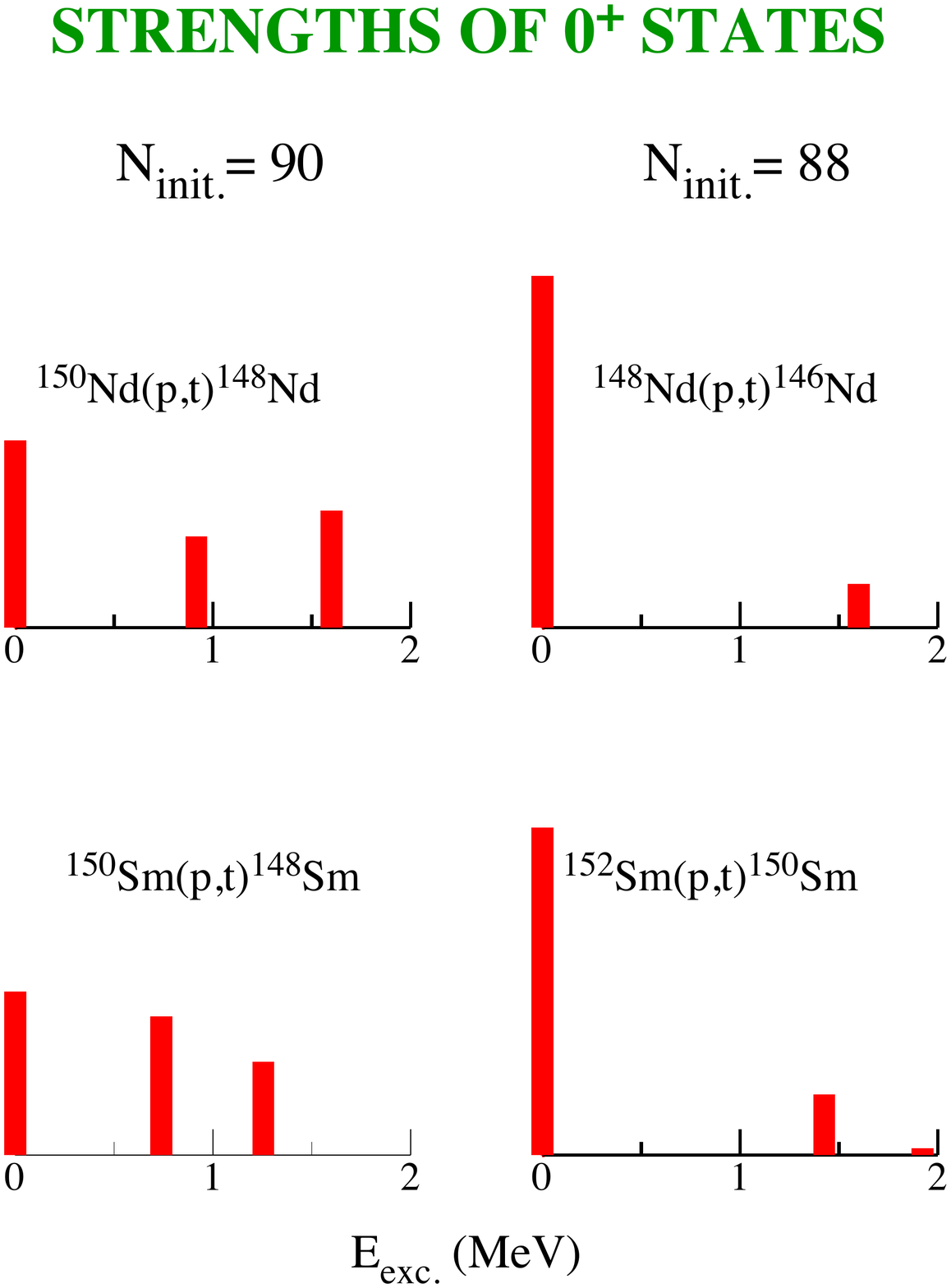}\end{center}

\noindent{\footnotesize {\bf Figure 9:} 
{The figure shows the relative cross section to 0$^+$ final states for the (p,t) reaction on Nd and and Sm targets with neutron numbers 88 and 90, normalized to the total strength in each case.  The removal of a pair of neutrons from N=90 nuclei displays serious fragmentation of the transition strength associated with the changing deformation in this step.  In the case of pair removal from N=88 nuclei, connecting two spherical states, the transition proceeds mostly from ground state to ground state.  The data plotted are taken from the references cited in \cite{ensdf}.}}
\vskip 0.5truecm 

The transitional nature of both parent and daughter nuclei in the $^{150}$Nd-Sm double beta decay system is likely to present 
more serious challenges for the calculations of the relevant matrix element.  Double beta decay involves the disappearance of  a  pair of neutrons from the parent ground state in a similar fashion to the removal of a neutron pair from the target ground state in the (p,t) reaction.  If the strength of the (p,t) reaction is altered by a factor of $\sim$3 by the presence of deformation, the question arises whether there are likely to be drastic changes in the double beta decay process as well.  

\subsubsection{$^{136}$Xe--$^{136}$Ba}
$\newline$ The nuclei associated with the $^{136}$Xe double beta decay nucleus are less well studied, largely due to problems associated with the production of targets. A measurement of the $^{138}$Ba(p,t) reaction leading to $^{136}$Ba  
 shows no evidence for a large cross section for an excited 0$^+$ state
 \cite{kusakari}. 

In proton-pair transfer several transitions were observed with significant strengths to 0$^+$ excited states on both $^{136}$Xe and $^{136}$Ba, similar to those on $^{130}$Te discussed above, and probably associated with the proton-pair vibration and a subshell at Z=64.  As for the case of $^{130}$Te, in this case there are also likely to be problems in the calculation of matrix elements within the QRPA formalism.

\section{Conclusions}

The discovery that neutrinos have finite rest mass has led to new
    interest in the search of neutrinoless double beta decay.  The developments of new large experiments to search for neutrinoless double beta decay with high sensitivity may have increased the probability that the observation of the process is on the horizon. The reliability of calculations of the associated nuclear matrix elements may soon become a rather crucial issue and should be tested in as many ways as possible.

In this article, two classes of experimental measurements are described where certain properties of the nuclear wave functions of parent and daughter nuclei are accessible, namely the microscopic decomposition of the occupancy of valence nucleon orbitals  and the extent to which the correlations between nucleons can be described as the pair correlations in the BCS approximation.   Both of these aspects would appear to have direct influence on the calculation of the matrix elements. 

The experimental methods and the present status of measurements were summarized, covering many of the candidates for this type of decay.  In at least one case where the valence occupancies were measured, calculations did not fit them and, when they were adjusted, significant changes in the matrix elements were found.  There are also several cases where the pair-adding and pair-removing strengths measured by transfer reactions are significantly fragmented and therefore not consistent with a simple BCS model. No adjustments for such effects have been made in the calculations.

To understand the implications of these results and to guide future measurements some further theoretical insights would be highly desirable.  Some of the questions, seen from the perspective of experimental measurements, are as follows:
\begin{itemize}

\item
 A qualitative understanding of the relative importance of the microscopic differences in the configuration of the components of the neutron pairs that decay into proton pairs, broken down orbit-by-orbit, would be highly desirable.  In other words, are the contributions to the matrix element relatively larger where the orbitals are the same ($e.g. ~g_{9/2}$ to $g_{9/2} ~vs.~  g_{9/2}$ to $p_{3/2}$)?  Are they larger for  components where a zero-node pair decays to a zero-node transition ($e.g. ~0f_{5/2}$ to $0g_{9/2} ~vs.~ 0f_{5/2}$ to $1p_{3/2}$)?  The answers to such questions are not clear, yet they are most likely buried in the details of calculations that have already been carried out.   Such insights could be important in guiding which measurements of occupancy are likely to be the most important.  (There are $some$ calculations in which histograms of contributions of matrix elements are shown \cite{vogel2}, but the relative importance also depends on the changes in occupancies.  Such information must be part of the same calculations, but a systematic study, separating the changes in occupancy, is not apparent in the literature.)

\item
To what extent do the BCS correlations matter?  We know that reactions that transfer a zero-coupled  pair of nucleons the cross sections are related rather simply to the pair-creation and pair-destruction operators of pairing (BCS) theory, and the coherence among the pairs is critical in determining the cross sections.  As was discussed, the BCS-like correlations between pairs of nucleons (that is assumed in QRPA calculations) is modified substantially in several of the candidate decays and this is reflected in the fragmentation of cross sections.  Is there a related reduction in the matrix elements?  

\item
Is the closure approximation valid?  If not, to what extent does it fail, and what aspects of the structure of the intermediate nucleus are relevant to this failure?  (Closure has been investigated and found to be a good approximation \cite{vogel1} within QRPA.  A more universal quantitative estimate of the validity of QRPA would be desirable.) 
\end{itemize}

Answers to these questions would be highly desirable because they would shed light on which experimentally accessible nuclear properties the calculations need to be able to reproduce in order to help constrain the calculations of matrix elements for neutrinoless double beta decay.

This work was carried supported by the U.K. Science and Technology Facilities Council and by the U.S. Department of Energy, Office of Nuclear Physics, under Contract No. DE-AC02-06CH11357. We would like to thank our many collaborators on these studies and the support from many individuals on the staff of the facilities at Yale, Osaka and Munich that were used to perform the experiments.

\end{document}